# Optoelectronic nibbling of laser linewidth using a Brillouin-assisted optical phase-locked loop

**GWENNAËL DANION,[1] MARC VALLET,[1], LUDOVIC FREIN,[1] PASCAL SZRIFTGISER,[2] AND MEHDI ALOUINI[1,*]**

[1]*Univ. Rennes, CNRS, Institut FOTON - UMR 6082, F-35000 Rennes, France*
[2]*Laboratoire de Physique des Lasers Atomes et Molécules, UMR 8523 Université de Lille—CNRS, 59655 Villeneuve d'Ascq, France*

**Corresponding author: mehdi.alouini@univ-rennes1.fr*

**We demonstrate that the implementation of phase-locked loop forbidding multimode operation of a long Brillouin resonator also leads to a dramatic reduction of the optical phase noise of the pump itself. In the case of a continuous Er,Yb:glass laser, a reduction by more than 90 dB at 100 Hz of the carrier is observed. This yields an optical linewidth estimated narrower than 2 Hz for the pump laser. The method being independent of the laser wavelength, it can be applied to almost any laser.**

***Key words:*** *Solid State Laser, Laser stabilization, Stimulated Brillouin Scatteri,.*

Narrow-linewidth lasers are required for many photonics applications, including, coherent optical communications systems [1], mm-wave and microwave generation [2], remote sensing based on coherent Lidar [3], high-resolution spectroscopy [4] and optical atomic clocks [5]. Standard laser linewidths typically range from 10 kHz for solid-state lasers up to a few MHz for semiconductors lasers. Several techniques have been developed in order to reduce the linewidth of a given laser to the sub-kHz level. Linewidth narrowing can for instance be achieved by servo-locking the mean laser frequency to a reference frequency provided by an atomic or molecular absorption line, leading to excellent long-term frequency stabilization. However, it does not generally enable the short-term phase-noise to be reduced. Other popular techniques consist in using the resonance frequencies provided by ultra-stable cavities [6] or km-imbalanced interferometers [7]. These methods are accurate but spurious cavity-length changes caused by environmental noise can degrade the laser linewidth. This leads to a tricky control of the environment of km-long delay lines or Fabry-Perot cavity. Frequency stabilization through injection locking provides alternative solutions for linewidth narrowing [8], but the problem is merely deferred as they rely on the use of stabilized low-noise optical seeders.

Brillouin fiber lasers (BFL) employing the stimulated Brillouin scattering in fibers [9] are known to present intrinsic narrow sub-kHz linewidths [10,11] and a low relative intensity noise [12]. Such an optical purity permits to target application such as high resolution spectral analysis [13] and microwave generation [14]. The spectral quality of BFL is mainly a consequence of the narrow homogeneous gain bandwidth, of the order of 20 MHz, which makes it possible to use significantly long fiber resonators. For instance, a typical 100-m-long fiber loop presents a free-spectral range FSR of about 2 MHz. Using a pump with a spectral linewidth smaller than 2 MHz can potentially ensure single mode operation. However, in order to circumvent mode-hopping of the generated Stokes wave, the pump wavelength has to be servo-controlled. One method, the so-called resonant pumping, consists in matching the pump wavelength with a loop resonance. This method implies rather complex optic and electronic stabilization schemes. It also implies the use of a narrow linewidth pump laser. Indeed, if the pump spectrum is broader than the cavity resonance width, part of the pump power is not injected into the cavity. A second method consists in using a non-resonant configuration for the pump in order to benefit from full pump power injection into the Brillouin resonator. However, this second configuration suffers from mode-hopping as well. To bypass this problem, we recently demonstrated that mode-hopping can be totally suppressed by means of an optical phase-lock loop (OPLL) locking the frequency difference between the pump and the Stokes waves to the frequency offset of the Brillouin gain [15]. Such a servo-loop ensures that the frequency of the Brillouin gain maximum is always tuned to the Stokes wave whose frequency follows in turn the resonator drift. This method has been shown to be very efficient for achieving robust single mode operation. In particular, it has been noticed that, in addition to the spectral narrowing, the power of generated Stokes wave increases by about 30% when the OPLL is closed. This apparent increase of the Brillouin efficiency raises the question of the influence of the PLL on the spectral properties of the pump itself.

The aim of this letter is thus to focus on the pump laser spectral linewidth. In particular, we will show that the OPLL which is initially implemented to ensure robust single mode operation plays a second role by increasing the spectral purity of the pump laser making its phase noise almost as low as that of the Stokes wave.

Figure 1 sketches the linewidth narrowing mechanisms behind non-resonant pumping of a Brillouin laser. Without OPLL (a) and assuming a fair pump linewidth, the BFL spectral purity is determined by the combination of the quality factor of BFL resonator and the Brillouin gain. Due to homogeneous broadening, BFL presents a single mode operation on a short time scale, as previously reported [13]. However, as the resonator length or the pump frequency drift, mode hopping might occur, which degrades the BFL spectral purity on longer time scales. Closing the OPLL (b) circumvents such mode-hopping. In addition, the pump-Stokes detuning being locked to an RF reference, the frequency noise of the pump line is expected to be correlated to that of the Stokes wave, leading to a transfer of the BFL purity to the pump. The combination of the BFL and the OPLL should then act as a linewidth narrowing module (LNM) for the pump laser.

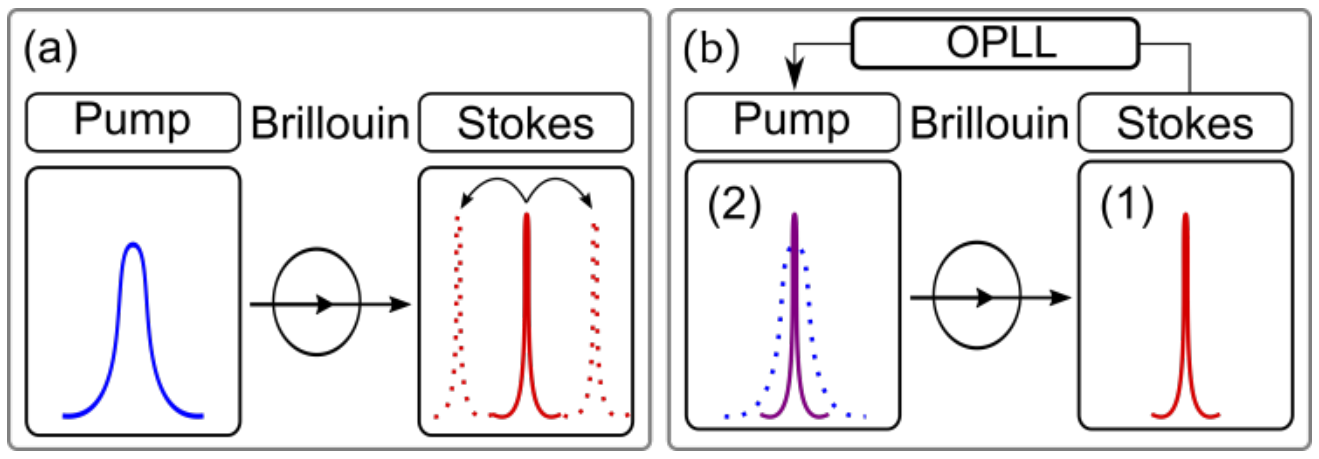

**Fig.ure 1.** Conceptual scheme of linewidth narrowing for the pump laser. (a) Without OPLL, the Stokes line presents mode-hops (dashed line) and the pump laser linewidth remains unchanged (b) with OPLL, the Stokes line is at first frozen. Once frozen, its spectral purity is expected to be reported to the pump, provided that the OPLL unity gain is larger than the pump linewidth.

Our setup is depicted in Fig. 2. The pump laser under consideration consists in a 6.8 mm long diode-pumped solid state laser. The active medium is a 1mm-thick codoped Er:Yb phosphate glass from Kigre Inc. The laser typically emits 6 mW at $\lambda_p = 1536$ nm when pumped with 400 mW at 980 nm. An intracavity etalon, made of a 100 µm thick $YVO_4$ crystal, ensures single-longitudinal-mode. A second crystal, consisting in a 500 µm thick transparent piezoelectric crystal of $Pb(Mg_{1/3}Nb_{2/3})O_3$–$PbTiO_3$ (PMN–PT) is inserted into the cavity. Due to the Kerr constant of PMN-PT, the laser frequency $\nu_P$ can be electrically tuned. We measured an electro-optic tuning coefficient $d\nu_P/dV$ equal to 20 MHz/V. The associated -3dB cut-off frequency was measured to be equal to 2 MHz.

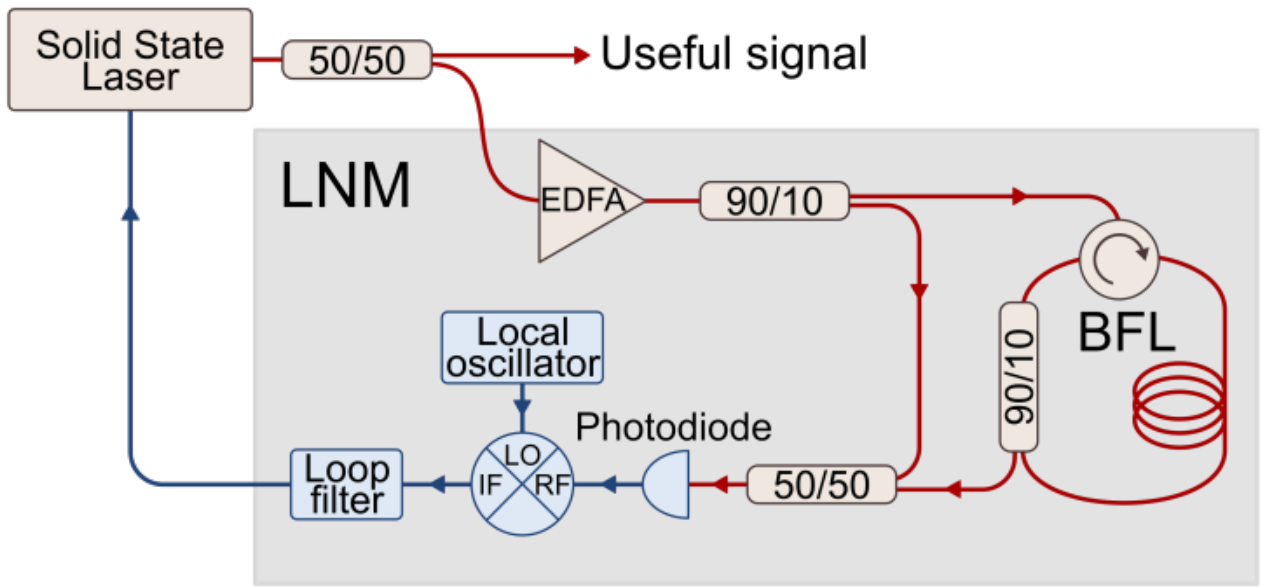

**Figure. 2.** Scheme of the Linewidth Narrowing Module. LNM : linewidth narrowing module, BFL: Brillouin fiber laser, LO: local oscillator, RF: radio frequency signal, IF: intermediate frequency. See text for additional details.

Part of the laser power (here 50%, i.e. 3 mW) is sent into the LNM. The optical beam is first amplified by means of a home-made Erbium-doped fiber amplifier, leading to a typical output power of 200 mW. 90% of the power is then used to pump the BFL. The architecture of the BFL is similar to that of Ref. [13]. The 110m-long ring cavity, made of polarization-maintaining optical fiber, contains an optical circulator for pump injection. It allows free propagation of the Stokes wave at frequency $\nu_S$ while blocking the pump beam of frequency $\nu_P$ after one round trip. Consequently this cavity is resonant for the Stokes wave but non resonant for the pump. An intracavity coupler permits to extract 10% of the Stokes wave. We point out that the whole BFL is molded in resin and thermalized to reduce acoustic and thermal fluctuations.

The typical 10 mW output Stokes power is recombined with the laser by means of a 2 × 1 coupler. The output is connected to a fibered photodiode (11-GHz bandwidth), with a trans-impedance gain of 440 V/W. The beatnote signal $\Delta\nu = \nu_P - \nu_S$ is then mixed with a local oscillator LO provided by an RF synthesizer. The output frequency was set at 10.998 GHz, corresponding to the frequency shift between the pump and the maximum of the generated Stokes band at 1536 nm. Finally the mixer output is directed to proportional-integrator active filter PI. The open loop unity gain bandwidth of the PI filter is 100 kHz, i.e., 10 times larger than the pump linewidth. It provides the pump with the error signal which is finally applied to the intracavity PMN-PT crystal in order to close the OPLL.

We have characterized the pump and Stokes optical linewidths using a self-heterodyning interferometer. As sketched in Fig. 3(a), one arm of the interferometer consists of a 700m-long optical fiber followed by Lefevre's loops for polarization control. A longer reference arm has been already used for characterizing Hz level laser linewidths where the Lorentzian lineshape of the heterodyne beatnote is recovered from the Voigt profile [16]. In our experiment the 700m long reference arm is chosen according to the expected pump laser linewidth in the free running regime, i.e. in the order of tens of kHz [17]. Any reduction of the extracted phase noise will prove the laser line narrowing effect that is pursued. The second arm contains an acousto-optic modulator driven at 80 MHz. The beating signal at 80 MHz is analyzed by an electrical spectrum analyzer.

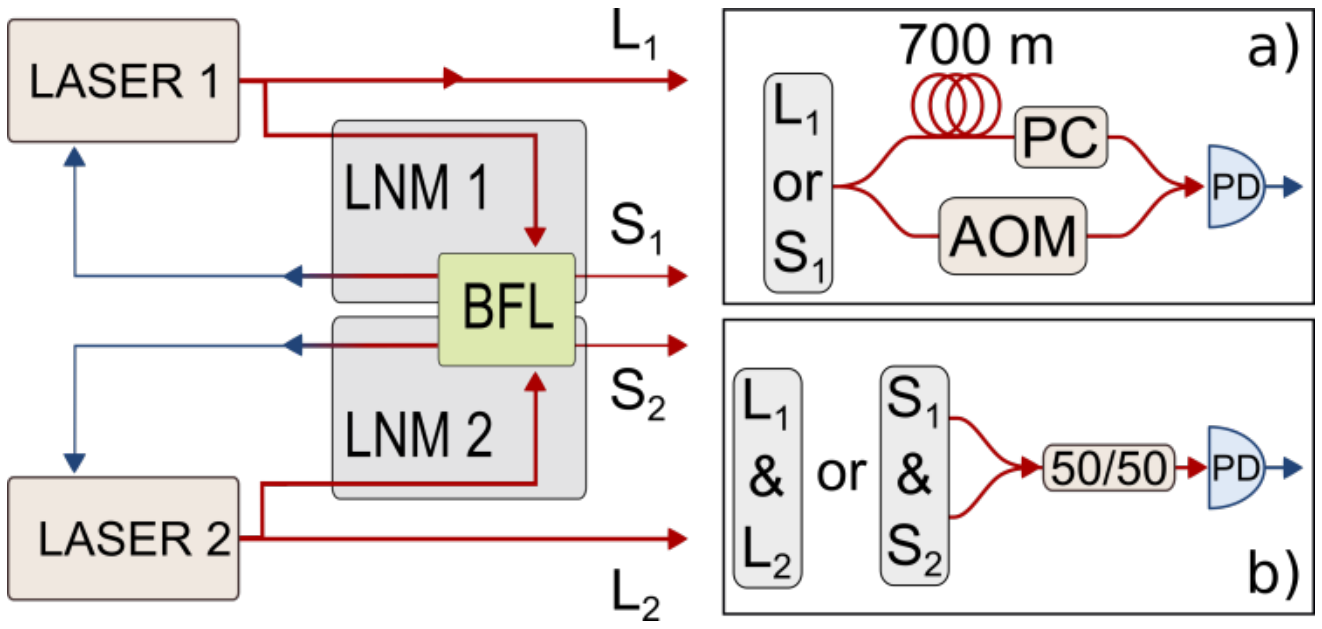

**Fig.ure 3.** Setup for characterizing the optical linewidths. $L_{1,2}$: laser. $S_{1,2}$: Stokes waves. (a) Self-heterodyning interferometer. (b) Beatnote between $L_1$ and $L_2$, or between $S_1$ and $S_2$.

First, the interferometer is fed by the laser while operating in the free running regime. The phase noise is deduced from the beatnote signal [14] and is reported on Fig. 4 (dark line). The phase noise power spectral density (PSD) is shown to follow $f^{-4}$, as expected for solid-state lasers. One possible evaluation of the optical linewidth $\Delta\nu$ (Full Width at Half Maximum) is provided by the following equation [18]:

$$\int_{\Delta\nu/2}^{\infty} S_\phi(f)\,df = 2/\pi \quad . \qquad (1)$$

According to the measured spectrum shape, the phase noise can be approximated by $10\log(A) - 10\,n\log(f)$ (in dBc/Hz), where $A$ is the phase noise offset and $n$ quantifies its slope, leading to an effective integral linewidth

$$\Delta \nu = 2\left(\frac{\pi A}{n-1}\right)^{\frac{1}{n-1}} \quad (2)$$

Equation 2 yields an estimated optical linewidth of 9 kHz. This linewidth is 200 times narrower than the 1.75-MHz FSR of the BFL cavity, thus ensuring single-mode operation of the Brillouin resonator despite mode hopping when the OPLL is not active.

When the OPLL is closed, the width of the spectrum associated to the beatnote at 80 MHz drastically decreases, as shown in the inset of Fig. 4. In fact, the phase noise PSD of the pump laser in closed loop operation presents a large decrease in the phase noise level (see Fig. 4, red line) with respect to the phase noise in the free running regime. In addition, it can be noticed that the phase noise level of the pump laser equals that of the BFL Stokes line up to 100 kHz (Fig. 4, green line) corresponding to the OPLL electrical bandwidth. This demonstrates that the LNM module effectively narrows down the pump laser linewidth by reporting the spectral quality of the BFL to the laser within the OPLL electrical bandwidth. It is important to mention that the pump phase noise remains squeezed as long as the OPLL is turned on, that is, for hours. Moreover, the PSD phase noise spectra hereby acquired in the closed loop configuration are instrument-limited. Indeed, the linewidth of the Stokes wave is expected to fall within the Hz level [15]. Measuring such a narrow linewidth would require increasing the length of the reference arm of the interferometer to around 10 km. Unfortunately, the use of such long delay line would degrade the measured phase noise because of its own optical length fluctuations. We thus choose to estimate the optical linewidth from the beatnote of two identical but independent Er:Yb lasers pumping the same BFL. For this purpose, a second tunable Er:Yb laser and a second OPLL have been built.

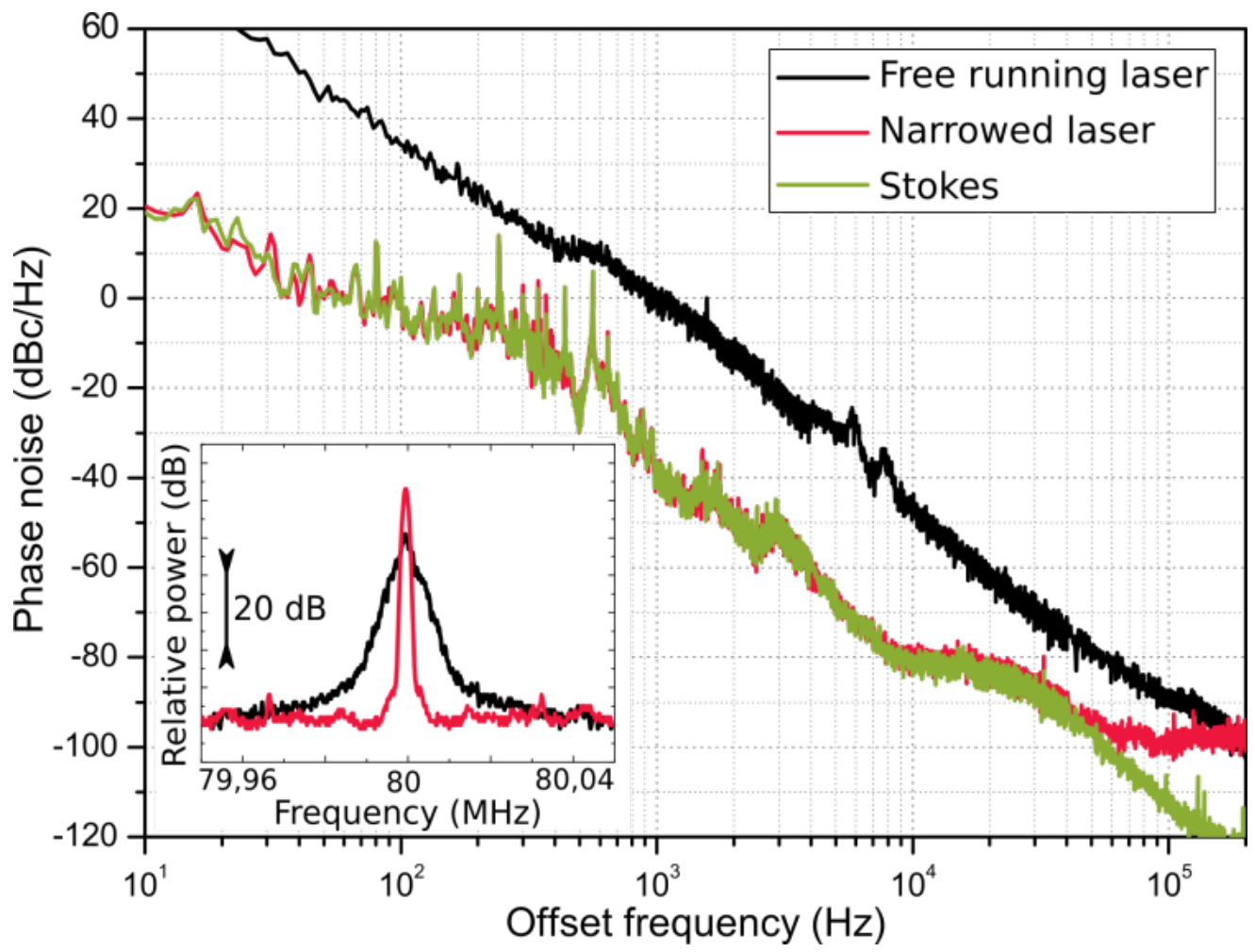

**Figure 4.** Optical phase noises. Inset: beatnote spectrum measured at the output port of the self-heterodyning bench; RBW 1 kHz; SPAN 100 kHz

As depicted in Fig. 3, the two lasers feed the same Brillouin resonator. The two generated Stokes waves sharing a single Brillouin resonator, they experience correlated fluctuations at low frequency. Thermal and acoustic noise impinging on the Brillouin resonator thus cancel out, providing a good indication of the short-term optical linewidth of the stabilized pump lasers. Lasers frequency difference $\nu_{L1} - \nu_{L2}$ was set to 20 GHz. Two independent OPLL, including two independent synthesizers and loop filters, are then used to lock $\Delta\nu_{1,2} = \nu_{L1,2} - \nu_{S1,2}$, at 10.998 GHz and 10.997 GHz respectively, corresponding to the Pump-Stokes frequency offset for $\nu_{L1}$ and $\nu_{L2}$.

By close analysis of the optical and electrical spectra for different pumping powers, we have verified that, even if they share the same resonator, the two Stokes waves, are not coupled through a nonlinear mechanism such as four-wave mixing for instance.

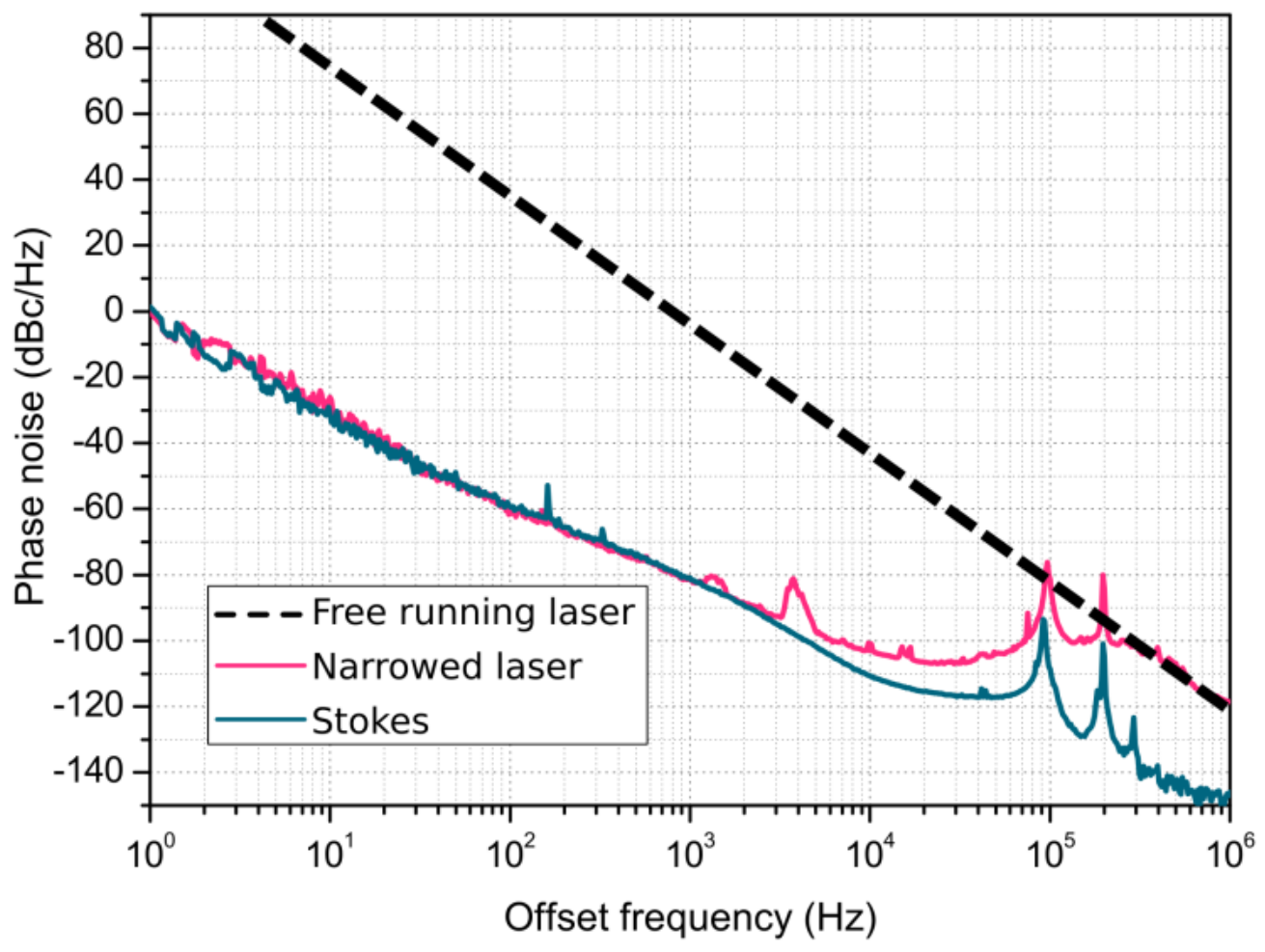

**Figure 5.** Phase noise of the beat note at 20 GHz between two stabilized lasers (blue), and between the associated Stokes waves (red). Dashed line: typical $f^{-4}$ optical phase noise of a free-running laser (as the one reported on Fig. 4).

Figure 5 reports the phase noise associated to the beatnote between such two independent stabilized pump lasers (see setup in Fig. 3(b)). One can see that compared to the optical phase noise of one laser, a reduction of more than 90 dB can be observed at 100 Hz from the carrier. Moreover, it is shown that the noise level of the beatnote between the two Stokes waves is remarkably well transferred to the pump lasers, within the loop bandwidth of 100 kHz. For frequencies out of the loop filter bandwidth, the noise level remains equal to that of the beatnote in free-running regime. In this condition, Eq. (2) yields an estimation of 3.2 Hz for the spectral width of the beating frequency. As the two lasers are independent and identical, we can thus assume that the optical linewidth of one pump laser is approximately half the beating frequency width, i.e., less than 2 Hz. Of course, the measured optical linewidth would have been probably lower if the two pump lasers had in addition, been stabilized to absolute optical references. Moreover, the cascade mechanism which takes place in the system may play a role in the narrowing process. More precisely, the Stokes linewidth is known to depend on the pump linewidth in BFL [19]. Consequently, the OPLL induced reduction of the pump linewidth will, entail an improvement of the BFL efficiency, which will, in turn, be reported back to the pump via the OPLL. To conclude on the existence of such cascaded narrowing effect, the intrinsic laser linewidth should be measured more accurately with a sub-Hz resolution apparatus [20].

In conclusion, we have shown that the proposed technique for getting rid of mode hopping in non-resonant pumped BFL is accompanied by a transfer of the Stokes wave spectral purity to the pump laser. This arrangement relying on an OPLL initially devoted to ensure robust single mode operation of the Brillouin resonator is actually an optoelectronic linewidth narrowing module. Indeed, it has been shown that the pump optical line is electronically nibbled until its width reaches that of the Stokes wave initially generated by the pump itself. This linewidth narrowing method can be applied to any laser as long as its frequency is electrically tunable and provided that the OPLL bandwidth is larger than the laser initial linewidth. For the sake of compactness, this method could benefit from microstructured fibers [21], new photonic on chip devices [22] or microcavity Brillouin resonators [23].

**Acknowledgment**. We thank Stéphanie Molin, Grégoire Pillet and Loïc Morvan for fruitful discussions, and Cyril Hamel for technical assistance. This project is partially funded by Agence Nationale de la Recherche (COM'TONIQ grant ANR-13-INFR-0011, 2014-2017). Contrat de Plan Etat-Région « Sophie/Photonique ».